\documentstyle[preprint,aps]{revtex}

\begin{document}
\draft
\preprint{Submited to Physical Review B}

\title{Equilibrium vortex-line configurations and critical currents in
thin films under a parallel field}  

\author{ Gilson Carneiro$^{*}$}
\address{Instituto de F\'{\i}sica\\Universidade Federal do Rio de Janeiro\\  
C.P. 68528\\ 21945-970, Rio de Janeiro-RJ \\ Brazil }
\date{\today}
\maketitle
\begin{abstract}

Vortex-line equilibrium configurations and critical currents for
type-II superconducting films at zero-temperature are studied
theoretically. The films are assumed to be of thickness less than or
equal to the 
penetration depth, free of pinning by imperfections, and
to be subjected to a  magnetic  field parallel to the film surfaces
and  to a transport current perpendicular to the field.
By numerical minimization of the exact London theory energy expression,
using simulated annealing techniques, the 
equilibrium configurations are determined with great accuracy over a
wide range of fields and currents. 
These consist of chains of straight vortex-lines whose number depends
both on the field and on the current. Transitions involving a change by
one in the chain  number are found to take place
either if the field is increased at zero transport current or if the
transport current is increased at constant field. At the transitions,
there is a considerable rearrangement in the vortex-line positions and a
small, discontinuous, change in their number. The equilibrium
configurations with   chain numbers that are not too small form  a
nearly perfect triangular lattice, centered with respect to the film
surfaces. However, small deviations from this arrangement are found to
be important in determining the behavior near the transition and when
the transport current approaches the critical current. It is found  
that the  critical current has a a non-monotonic
dependence on the field. The zero-temperature equilibrium phase diagram
in the field-transport current plane is reported.

\end{abstract}

\narrowtext

\section{Introduction} 
\label{sec.int}

There is a great deal of activity nowadays in the investigation of the
properties of the mixed-state in type-II superconductors under various
physical situations. 
One problem of interest is  the study of vortices in 
films of thickness comparable with the penetration depth placed on an
external magnetic field  parallel to the film  surfaces.
In this case vortices enter the film as straight lines parallel to
the field direction. The shielding current penetrate the full
thickness of the film and pins the vortex-lines to the film's interior.
A transport 
current applied parallel to the film surfaces and perpendicular to the
vortex-lines will not dissipate energy unless its magnitude is
large enough for the Lorentz force exerted by it on the
vortex-lines  to overcome the force generated by the
shielding currents and by the interactions between the vortex-lines.
Therefore, even in the absence of pinning by imperfections, the mixed
state in the film  can support a finite transport current without
dissipating energy, that is in thermodynamic equilibrium.

Previous theoretical work on this problem has calculated, in the London
limit, the equilibrium  vortex-line configurations and critical
currents using both analytical  \cite{car,shm,rumk,my,tak}  and numerical
\cite{bron} methods . These find that 
the equilibrium configurations consist of vortex-line chains and that
there are transitions, induced by changing the applied magnetic field,
in which the number of chains changes.

Measurements on films in this geometry of the
magnetization and of the critical current have been reported by
several authors \cite{bron,gui,zie,yam,fog,lob,ste}. It is found  that the
magnetization has  peaks as a function of the field, not present in
the bulk material, which are interpreted as resulting from the
transitions involving the change in the number of 
chains \cite{bron,gui,zie}. Some of the critical current measurements 
find non-monotonic dependence on the field that is
also attributed to these transitions \cite{yam,fog,lob}.
Other measurements find no such dependence, but the large critical
currents observed are interpreted as due to pinning of the vortex-lines
by the shielding current \cite{ste}.
 
The aim of this paper is to calculate in detail the vortex-lines
equilibrium configurations and critical currents at zero temperature in
films of thickness less than or equal to the penetration depth, without
pinning  by imperfections, in the presence of  both a parallel
field and  a transport current. The starting point of this
calculation is the known expression for the energy of a system of
straight vortex-lines parallel to the film surfaces, exact in the
London limit  for sufficiently thin films \cite{car,shm,rumk,my,tak,bron}.
Minimization of this energy is carried out numerically, using simulated
annealing techniques, assuming only that the distributions of vortex 
lines in the film is periodic. By this method  the equilibrium
vortex-line configurations are determined  very accurately over a wide
range of external fields and transport currents. Several new results
are reported. 

The equilibrium configurations obtained by this method are found to
consist of vortex-lines chains, all  with the same
intra-chain spacing between vortex-lines.
In the absence of a transport current, it is found 
that,  as the external field grows above the  lower critical
field, there are regions where the number of chains remains constant in
which the  number of vortex-lines in the film grows smoothly with the
field. Transitions consisting in a change by one of the number of
chains   are found to take place at  critical field values.
These results are consistent with previous ones, obtained 
using a similar numerical method \cite{bron}. The  present calculation
reveals that at  the transitions, besides the considerable
rearrangement in the vortex-lines positions found previously 
\cite{bron},  their number has a small discontinuous increase.  
Even for a modest number of chains, the equilibrium 
vortex-line configurations  are found to be close to a triangular
lattice, as predicted in Ref. \cite{shm,rumk} . However, it is found that
there are  significant  differences between the triangular lattice and the
equilibrium  configurations, not reported previously,  that play an
important role in the vicinity of the above described transitions.
The changes in the equilibrium vortex-line configurations
caused by a  transport current are studied in detail and the critical
current is calculated. These results are also new. It is 
found that, by changing the transport current at 
constant field up to the critical current, two distinct behaviors take
place, depending on the field value. In the first the number of chains
remains the same as that for no current. The  effects of the
current are  to shift the chains towards the film surface and to
increase a little the number of vortex-lines. Near the critical current
this shift is non-uniform, contrary to  previous calculations that find
an uniform shift \cite{shm}. It is found that, if the vortex
chains were forced to shift uniformly, a critical current
considerable larger than that obtained from the exact calculation would
result. 
The second type of behavior is one in which, as the current
grows past a certain value, smaller than the critical current, a
transition to a configuration with one additional chain takes place.
This is similar to the transition that occurs by increasing the field
without a transport current and has not been predicted in previous
calculations. 
The critical current versus field curve is found to be non-monotonic with
a structure that is a direct consequence of the vortex-line chains
configurations that exist in the film. The zero-temperature equilibrium phase
diagram  in the external field-transport current plane is obtained.

This paper is organized as follows. In Sec.\ \ref{sec.ldt}. the
London theory results for the energy are briefly reviewed. In Sec.\
\ref{sec.emin} the numerical method for energy 
minimization is presented. In Sec.\ \ref{sec.res} the results obtained
by this method are reported and their physical significance is
discussed. Finally, in Sec.\ \ref{sec.con}  the predictions of this
paper are compared to experiment and its conclusions stated.

\section{ London theory}
\label{sec.ldt}

Consider a film made of a type-II superconducting material of thickness
$D$, length $L$ and width $L$. The material is assumed to be
uniaxially anisotropic, with anisotropy axis perpendicular to the film
surfaces (the c-axis) and
characterized by anisotropy parameter $\gamma$ 
($\equiv \sqrt{m_{ab}/m_{c}})$. The other material 
parameters are the penetration 
depth $\lambda$, for currents flowing parallel to the a-b plane, and
the coherence lengths in the a-b plane $\xi$. 

The film is assumed to be subjected to an external field $H$ parallel
to the film surfaces (along the b-direction) and to
a transport current flowing perpendicular to $H$ and also parallel to the
film surfaces (a-direction), whose average current
density is $J_t$ ( Fig.\ \ref {fig.uc}).  

The interest here is to obtain the equilibrium vortex-line
configurations and the critical currents at zero temperature for
$D\leq \lambda$, in the London limit, $\lambda \gg \xi$. It is 
assumed that: i) the vortex-lines are straight and parallel to the field
direction (b-direction). ii) Their distribution in the ac-plane 
is periodic in the $a-$ direction, with period $a_v$.  This periodic
structure  consists of $L/a_v$ identical primitive unit cells
with dimensions $D\times a_v$ placed side by side. Each  cell contains 
$n_v$  vortex-lines located at $(x_{aj},x_{cj})$, $j=1,2,...,n_v$, with 
$0\le x_{aj}< a_v$ and $0\le x_{cj}\le D$ ( Fig.\ \ref {fig.uc}). Even
though these assumptions allow for very general vortex-line
configurations, the  
 calculations reported here find that the equilibrium ones  consist of 
$n_v$  chains of vortex-lines located at $\{x_{cj}\}$, with vortex
lines uniformly spaced from one another within each chain by $a_v$
( Fig.\ \ref {fig.uc}).

The solutions of London equations for the film with an arbitrary
distribution of straight vortex-lines  parallel to the film surface
were obtained by several authors using the method of
images \cite{car,shm,rumk,my,tak,bron}. One such a vortex-line, with
vorticity $q=1$ and located at 
$(x_a,x_c)$ ($D\geq x_c \geq 0$), generates two infinite sets of mirror
images. One set has images with vorticity $-q$ located at  $-x_c +
2nD$, $n=0,\pm 1, \pm2, ...$. The other has images with vorticity
$q$ located at  $x_c + 2mD$, $m=\pm 1, \pm2, ...$. 

The total energy  of an arbitrary distribution of such vortex-lines can
be written as
\begin{equation}
E=E_{int}+ E_{self}+E_{H}+E_{J} \; .
\label{eq.et}
\end{equation}
In Eq.\ (\ref {eq.et}) $E_{int}$ is the energy of interaction of
vortex-lines with one 
another and with the images, excluding the self-images; $E_{self}$ is the
vortex-lines self energy plus the energy of interaction with their own
images; $E_{H}$ is the energy of
interaction of the vortex-lines with the screening current  generated
by the external field and $E_{J}$ is the energy of
interaction of the vortex-lines with the transport current.

For an isotropic film ($\gamma=1$), and in the limit where $(\pi
\lambda/D)^2 \gg 1$, $E_{int}$ and $E_{self}$ are 
given by
\begin{equation}
E_{int}/L^2= \epsilon\frac{1}{2a_v}\sum_{n,i,j}
\ln{[\frac{\cosh{\pi(na_v+x_{ai}-x_{aj})/D} - \cos{\pi(x_{ci}+x_{cj})/D}}
{\cosh{\pi(na_v+x_{ai}-x_{aj})/D} - \cos{\pi(x_{ci}-x_{cj})/D}}]} \; ,  
\label{eq.eint}
\end{equation} 
where  $\epsilon=(\phi_0/4\pi \lambda)^2$ and
$\sum_{n,i,j}$ runs over $n=0,\pm 1,...,\pm
L/a_v $, and $i,j=1,2,...,n_v$, excluding for $n=0$ the $i=j$ term ;    
\begin{equation}
E_{self}/L^2 =\epsilon\frac{1}{2a_v}\sum^{n_v}_{j=1}
\ln{[\frac{4\sin^2{(\pi x_{cj}/D)+(\pi\xi/D)^2}}
{(\pi\xi/D)^2}]} \; ,  
\label{eq.eself}
\end{equation}

The energies of interaction with the currents are given by
\begin{equation}
E_{H}/L^2 =\epsilon \frac{4\pi \lambda^2H}{\phi_0 a_v}\sum^{n_v}_{j=1}
 [\frac{\cosh{(x_{cj}-D/2)/\lambda}}{\cosh{D/2\lambda}} -1] \; ,  
\label{eq.eh}
\end{equation} 
and
\begin{equation}
E_{J}/L^2 =\epsilon\frac{8\pi^2 \lambda^2 DJ_t}{\phi_0 a_v c}\sum^{n_v}_{j=1}
[\frac{\sinh{(x_{cj}-D/2)/\lambda}}{\sinh{D/2\lambda}}+1] \; ,  
\label{eq.ej}
\end{equation}

The total energy for an anisotropic film ($\gamma \neq 1$) is
related to that for an isotropic film ($\gamma = 1$), with the same
$\lambda$ and $D$, by the following scaling relation \cite{bla}
\begin{equation}
E[\xi, a_v,\{(x_{aj}, x_{cj})\},H,J_t;\gamma] =\gamma  
E[\gamma \xi, \gamma a_v,\{(\gamma
x_{aj},x_{cj})\},\gamma^{-1} H,\gamma^{-1}{J_t} 
;\gamma=1] 
 \; .    \label{eq.scal}
\end{equation}  
Thus the equilibrium configurations  for
an anisotropic superconductor follow  from the corresponding
ones for the isotropic material by scaling the parameters according to
Eq.\ (\ref {eq.scal}). In what follows the
discussions will be limited to isotropic materials.

In order to obtain the equilibrium vortex-line configuration it is
necessary to find the minimum of $E$, Eq.\ (\ref {eq.et}), for  given
$H$ and $J_t$. This is discused next.

\section{Energy minimization}
\label{sec.emin}

The energy minimization is carried out numerically using  simulated
annealing. For a given $H$ and
$J_t=0$, the following procedure is used. Assuming that there are $n_v$ 
vortex-lines within the unit cell, a standard Monte Carlo (MC)
procedure that moves vortex-lines within a single cell is used 
to find the  positions $(x_{aj},x_{cj})$ and cell length
$a_v$ at a given "temperature"  \cite{sma}.  This fictitious temperature is used to
overcome energy barriers.  By lowering it the minimum of $E/L^2$, Eq.\
(\ref {eq.et}), is found. The equilibrium configurations are obtained
from this data by 
comparing the minimum energies for different $n_v$ and  determining the
lowest one. 

For $J_t\neq 0$ it is necessary to search for a local minimum of
$E/L^2$, Eq.\ (\ref {eq.et}), at fixed $n_v$. The reason is that, for
fixed $n_v$ and for $J_t>0$,  $E$ is unbound because vortex-lines
placed at $x_{c}=D$  have zero 
interaction energy and self-energy, but negative $E_{J}$. Thus the energy
can always be lowered by putting more and more vortex-lines at
$x_{c}=D$ close together. The local minimum is, clearly, the physical
relevant one. 

For given $H$ and $n_v$ the local minimum is determined as follows.
Starting from the vortex-line configuration found for these values of $H$,
$n_v$ and for $J_t=0$, a new configuration that minimize $E$ is determined
for small $J_t$  by  running the MC procedure at very low
"temperature". This  avoids "thermal" fluctuations that might
overcome the energy barrier separating  the local minimum from
the unphysical configurations  described above. By further increasing
$J_t$ by small amounts new local minima configurations are found.
The equilibrium configurations for $J_t > 0$ are determined as in the
case where $J_t=0$, described above, by comparing the energies of the
local minima for different $n_v$. It is found that 
above a certain value of $J_t$ the vortex-lines accumulate at $x_c=D$,
their number growing without limit. This mimics what happens above the
critical current, where vortex-lines enter the film at $x_c=0$, are
dragged by the transport current to the film surface at $x_c=D$, where
they are annihilated. The smallest value of of $J_t$ where this happens
is interpreted as the critical current $J_c$. 

This method allows the equilibrium configurations of interest 
to be determined accurately using small values of $n_v$ and  with modest
personal computers. 

In the results reported in  Sec.\ \ref{sec.res} $\lambda$ and $\xi$ are
considered as fixed lengths. Films with $\xi=10^{-2}\lambda$
and $D=\lambda,\; \lambda/2,\; \lambda/4$  are investigated over a
wide range of values of $H$ and $J_t$. According to Eq.\ (\ref
{eq.et}), for these values of $D$, $E$ is weakly dependent on
$\lambda$. The mathematical derivation of Eq.\ (\ref {eq.eint})
requires only that $(\pi \lambda/D)^2\gg1$, which is satisfied.

The natural units for the external field and for the transport current
are the lower critical field, $H_{c1}$, 
and of the depairing current, $J_d$, respectively. For the films with
$D\leq \lambda$  these quantities are given by
\begin{equation}
H_{c1}=\frac{\phi_0}{\lambda^2}
\frac{\ln{(2D/\pi \xi)}}{4\pi(1-1/\cosh{D/2\lambda})} 
\; ,    \label{eq.hc1}
\end{equation}
and
\begin{equation}
J_d=\frac{1}{12\sqrt{3}\pi^2}\;\frac{c\phi_0}{\xi\lambda^2} 
\; .   
\label{eq.jd}
\end{equation}

For the films studied here the values of $H_{c1}$ in units of
$\phi_0/\lambda^2$ are: \linebreak
$H_{c1}(D=\lambda)=2.92, \;H_{c1}(D=\lambda/2)=9.04$, and
$H_{c1}(D=\lambda/4)=28.37$.

\section{Results and Discussion}
\label{sec.res}

For $J_t=0$, and for $H$ just above $H_{c1}$,  the equilibrium
configuration is a single vortex-chain ($n_v=1$) located at the film
center ( Figs.\ \ref {fig.vcl} and \ \ref {fig.chp}). As $H$ increases
the vortex-lines  
in the chain come close together, that is $a_v$ decreases, as shown in
 Fig.\ \ref {fig.avh}. Eventually a value of $H$ is reached where the
single chain 
splits in  two chains ($n_v=2$), located symmetrically about the film
center ( Figs.\ \ref {fig.vcl} and \ \ref {fig.chp}). This two-chain
configuration exists over a range of $H$ 
values, above which the energy minimum corresponds to three chains
($n_v=3$) ( Figs.\ \ref {fig.vcl} and \ \ref {fig.chp}.). As $H$ is
further increased  the pattern is as follows. A stable $n_v$-chain
configuration occurs over a range of $H$, followed by a  $n_v+1$-chain
configuration as $H$ grows past  a critical value. At the
$n_v\rightarrow n_v+1$ -transition $a_v$ increases discontinuously, with
the amplitude of the discontinuity decreasing with increasing $n_v$ ( or
$H$), as shown in  Fig.\ \ref {fig.avh}. Similar results are obtained
for other $D$ values.  

In  Fig.\ \ref {fig.nah} the total number of vortex-lines per unit
length of the film $n_v/a_v$ is shown as a function of $H$. It is seen that
the number of vortex-lines increases roughly linearly with $H$, with
constant slope 
in the regions where $n_v$ is constant. At the 
$n_v\rightarrow n_v+1$ chains transition there is  a discontinuous jump
in $n_v/a_v$, by about $10\%$ in the region covered by  Fig.\ \ref
{fig.nah}. The relative size  of the discontinuities 
in both $a_v$ and $n_v$ are considerably larger ( Fig.\ \ref
{fig.avh}). However,  these quantities increase by about the same
factor so that the relative increase in their ratio  is smaller.
At higher values of 
$H/H_{c1}$ than those shown in  Fig.\ \ref {fig.nah}  the number of
vortex-lines continues to increase
in the same fashion,  but the discontinuity at the transitions become
smaller.

Even for modest values of $n_v$, the equilibrium configurations are
found to be very 
close to a perfect  triangular lattice centered with respect to the
film's surfaces ( Fig.\ \ref {fig.tri}), as predicted in Refs.
\cite{shm,rumk} . However, the vortex 
line configurations obtained by numerical minimization of Eq.\ (\ref
{eq.et})  show important differences with this triangular lattice, as
discussed next.

In the centered triangular lattice the
distance between adjacent chains is $d_c=D/(n_v+1)$, which is also the
distance from the first and last chains to the film surfaces at $x_c=0$
and $x_c=D$, respectively, and $a_v=d_c/\sin{60}$ ( Fig.\ \ref
{fig.tri}). The results of the present calculation show that  the chain
spacing 
increases slightly towards the film center. Numerical minimization of
$E$, Eq.\ (\ref {eq.et}), assuming a centered triangular lattice
structure, gives  the 
$a_v\times H$ curve with steps shown in  Fig.\ \ref {fig.avh}. In this
approximation  $a_v\simeq D/(n_v \sin{60})$,  and $n_v$  depends on $H$
through the 
minimization of $E/L^2$. Consequently, $a_v$ is independent of $H$ in the
range  where  $n_v$ is constant, in sharp disagreement with the results
described above. However, it is apparent from  Fig.\ \ref {fig.avh}
that the triangular 
lattice prediction agrees well with the mean value of $a_v$ in the
range where $n_v$ is constant. At the $n_v\rightarrow n_v+1$
chains transition $d_c$ and $a_v$ decrease
discontinuously to fit a centered  triangular lattice with
$n_v+1$ chains. This leads to a  
$(n_v/a_v)\times H$ curve with steps: $n_v/a_v$ is constant in the range
where both $n_v$ and $a_v$ are constant and increases discontinuously by 
$\Delta (n_v/a_v)\simeq 2n_v\sin{60}/D$ at the transition. This 
 linear dependence on $n_v$ of $\Delta (n_v/a_v)$ is also in
strong disagreement with the results of the full minimization of Eq.\
(\ref {eq.et}), discussed above. 

A simpler way to estimate $n_v$ and $a_v$ for large $n_v$ is as follows.
First replace
the discrete vortex distribution by a continuous one
$\nu_v(x_c)$(=number of vortex-lines per unit area).
Minimization of Eq.\ (\ref {eq.et}) under this assumption gives
\begin{equation}
\nu_c(x_c)=\frac{H}{\phi_0}
\frac{\cosh{(x_c-D/2)/\lambda}}{\cosh{D/2\lambda}} 
\; .    \label{eq.nuc}
\end{equation}
This result shows that the vortex-line density is slightly larger near
the film's surfaces, in accordance with the results of the full
minimization of Eq.\ (\ref {eq.et}), as discussed above. Next
approximate the average vortex-line density, $n_v/a_vD$, by the average
over the film thickness of $\nu_c(x_3)$ which, according to Eq.\ (\ref
{eq.nuc}) is  $\bar{\nu_c}=(H/\phi_0)(2\lambda/D)\tanh{(D/2\lambda)}$,
and use for 
$a_v$ the centered triangular lattice approximation discussed above.
For $n_v\gg 1$ this gives $n_v=D/(a_v\sin{60})$ and
\begin{equation}
a_v=
[\frac{2\lambda H \sin{60}}{D\phi_0}\tanh{\frac{D}{2\lambda}}]^{-1/2} 
\; .    \label{eq.atr}
\end{equation}
This prediction agrees well with the average of $a_v$ in the interval
where $n_v$ is constant, even for  modest values of $n_v$, as show in
 Fig.\ \ref {fig.avh}. It shows that this average value decreases with
$H^{-1/2}$. However, this approximation overestimates $n_v$ in the
range shown in  Fig.\ \ref {fig.avh}.

For $D\leq \lambda$ the magnetic induction 
\begin{equation}
B =\frac{\phi_0 }{a_vD}\sum^{n_v}_{j=1}
 [1-\frac{\cosh{(x_{cj}-D/2)/\lambda}}{\cosh{D/2\lambda}} ] \; ,
\label{eq.ind}
\end{equation} 
resulting from the vortex-line configurations in the field is small.
It follows from Eq.\ (\ref {eq.ind}) that $B\sim
\phi_0(n_v/a_vD)(D/\lambda)^2/12$. 
Consequently, $M=(B-H)/4\pi$  deviate very little from the Meissner
state value $M=-H/4\pi$.

For $J_t> 0$, and if $H$ is not too close to the field where the
$n_v\rightarrow n_v+1$ chains transition takes place at $J_t=0$, 
the equilibrium vortex configuration has  the same $n_v$ as that for
$J_t = 0$ and the same $H$. The chains  positions are shifted
towards the film surface located at $x_c=D$,  $a_v$ 
decreases slightly  and, consequently, the number of vortex-lines in
the film increases a little. The shift
grows with $J_t$  and is non-uniform, with the chain closest
to film surface  at $x_c=D$ shifting the most, as shown in  Fig.\ \ref
{fig.chpj} .
In this figure it is seen that as $J_t\rightarrow J_c$  the chain
closest to the $x_c=D$ film surface  is about halfway between its
position at $J_t=0$ and  this surface.  

A simple model to calculate $J_c$, proposed in Ref. \cite{shm} ,  is to
assume that the vortex-lines form 
a rigid triangular lattice. In this case the effect of the force
exerted by the transport current on the vortex-lines is to
displace the lattice uniformly towards the film surface. Equilibrium is
reached when this force  is balanced by that  exerted by the other
vortices and by the shielding current.  Numerical calculations based on
this model and on Eq.\ (\ref {eq.et}) show that above a critical value
of $J_t$ the rigid lattice has no 
equilibrium position within the film. However, this critical value
is found to be considerably larger than that calculated as
previously described. This shows that, at least for the range of $H$
values studied here, the subtle changes  in the vortex-lines
configuration away from the triangular
lattice  caused by  $J_t$ play an important role in determining
the critical current.

If $H$ is close to the field for the $n_v \rightarrow n_v+1$ chains
transition  at $J_t=0$, it is found that a similar transition takes place
by increasing  $J_t$ at constant $H$.

The vortex-line structure in the film can be summ
arized  in a
zero-temperature equilibrium phase diagram, as shown in  Fig.\ \ref
{fig.phd}.  
For $D=\lambda/2$ and $D=\lambda/4$ the phase diagram is similar.
The critical currents for these $D$ values  are show  as a function of
$H$ in  Fig.\ \ref {fig.jch}.

\section{Conclusions}
\label{sec.con}

The calculations reported  in  Sec.\ \ref{sec.res} describe
the behavior of an ideal model. That is, vortex-lines in equilibrium at
zero temperature in films without imperfections under
a field that is precisely parallel to the film
surfaces. Even a very small angle between the field and the film
surfaces, which is always present in experiments, invalidates these
predictions, since dissipation occurs when $J_t\neq 0$. In such fields 
the vortex-lines are tilted. The Lorentz force exerted by $J_t$ on
these lines has a component along the b-direction which will move the
lines and dissipate energy. However, since the tilt is small, it takes
only a small amount of pinning to prevent this motion. Morever, it is
reasonable to assume 
that the  tilted vortex-line configurations are very close to  those 
for untilted vortex-lines at the same field. That is, the vortex-lines
are nearly parallel to the b-direction and their average positions in
the ac-plane coincides with those obtained for lines parallel to the
b-direction. Thus, the ideal model predictions are applicable to
equilibrium properties of films, provided only that pinning by
imperfections  is sufficiently weak and that the field makes a small
angle with the film surfaces.     

It has been usual to relate  peaks in the magnetization data
on films with the
$n_v  \rightarrow n_v+1$  chains transitions that occur in the
ideal model  \cite{bron,gui,zie}. It is therefore interesting to discuss
what are the predictions of  the ideal model  to the behavior of the
magnetization. 

As discussed in Sec.\ \ref{sec.res}, the magnetization due to
vortex-lines  that are 
exactly parallel to the film surfaces deviates very little from the
Meissner value.

If the vortex-lines are tilted with respect to the b-direction the 
equilibrium magnetic induction  has a c-component $B_c$. Assuming that
the  tilt is uniform, $B_c$  is proportional to the
number of vortex lines per unit area, that is 
 $B_c \propto \phi_0(n_v/a_vD)$. The equilibrium magnetization has also
a c-component  $M_c\propto \phi_0(n_v/a_vD - H/\phi_0)$. Thus, $M_c$ 
is a direct measure of number of vortex-lines in the film. 
In  Fig.\ \ref {fig.nah} (inset) $m_c=-(n_v/a_vD- H/\phi_0)\lambda^2$ is
plotted versus $H$.  The peaks seen in this figure result from the
increase in the number of vortex-lines in the film that take place at
the $1 \rightarrow 2$  and $2 \rightarrow 3$ chains transitions.
Similar results are obtained for larger values of 
$D$. Experiments on Nb/Cu multilayers reported in Refs. \cite{bron,gui,zie}
find that  the $M_c\times H$ curve has peaks  at  field values
approximately equal to those where the $n_v
\rightarrow n_v+1$ chains transition takes place in the ideal model. If
these experiments measure the equilibrium $M_c$, there  is the
following disagreement 
with the ideal model predictions. The  experimental  $\mid M_c \mid$ decreases
in the range of $H$  where $n_v$ is constant, whereas the 
ideal model predicts that it increases. If in the experimental setup
the vortex-lines in the film are not in equilibrium, then it is unclear
how the $M_c$ peaks are related to a transition that only takes place
in equilibrium.

In the experiments reported in Refs.  \cite{bron,gui,zie}  $M_c$ is
measured  as $H$ is 
changed at $J_t=0$. According to the results reported above the $n_v
\rightarrow n_v+1$ chains transition can also take place if $J_t$ is
increased at constant $H$. This suggests that, if peaks in $M_c$ are
indeed related to the $n_v\rightarrow n_v+1$ chains transitions, 
it may be possible to observer peaks in $M_c$ by increasing $J_t$  at
constant $H$.   

The measurements of the critical current in  films
under a parallel field reported in Refs. \cite{yam,fog,lob,ste} can also be
compared to the ideal model predictions. 
The main characteristics of these predictions are a
non-monotonic $J_c\times H $ dependence and  values of $J_c\sim 
1-9\times 10^{-2}J_d$. The results  reported
in Refs. \cite{yam,fog,lob} show non-monotonic dependence on $H$, 
with a characteristic features at $H$ values comparable to those where 
$n_v\rightarrow n_v+1$-chains transition take place. However the critical
currents measured in Refs. \cite{yam,lob}  are one order of magnitude
smaller than those predicted by the present calculation.
The results reported in Ref. \cite{ste} show only a monotonic
dependence on $H$,  and find $J_c$ values that are about one order of
magnitude larger than those shown in  Figs.\ \ref {fig.phd} and \ \ref
{fig.jch}.   

In conclusion then, the ideal model predictions are not fully supported
by the experimental data. However, this model is useful as a  starting
point to study the interesting behavior of films with weak pinning
under parallel fields. Further work is necessary in order to
incorporate into the model the effects that are neglected here.

\acknowledgments
It is a pleasure to thank Prof. M. Doria for stimulating  conversations.
This work was supported in part by FINEP/Brazil,
CNPq-Bras\'{\i}lia/Brazil and Funda\c{c}\~ao Universit\'aria Jos\'e
Bonif\'acio.

\begin{figure}
\caption{A portion of the film cross-section showing 
a typical vortex-line configuration. The primitive unit
cell is indicated by the double arrow. The field, $H$, is
entering the paper ( b-direction). The transport current direction is
indicated by the single arrow.}
\label{fig.uc}
\end{figure}

\begin{figure}
\caption{Equilibrium configurations for $J_t=0$ and $D=\lambda$. The
field values are: $H=2H_{c1}$ for $n_v=1$; $H=3H_{c1}$ for $n_v=2$; 
$H=5H_{c1}$ for $n_v=3$; $H=7.25H_{c1}$ for $n_v=4$; 
$H=11H_{c1}$ for $n_v=5$ and $H=15H_{c1}$ for $n_v=6$.   }
\label{fig.vcl}
\end{figure}

\begin{figure}
\caption{Vortex-chain positions for $J_t=0$ and $D=\lambda$. The
continuous lines are just guides to the eye.  }
\label{fig.chp}
\end{figure}

\begin{figure}
\caption{Intra-chain vortex-line spacing for $J_t=0$ and $D=\lambda$.
Thick line:  full minimization of $E/L^2$ .
Dashed curve: continuous approximation. Inset: 
comparison with the centered triangular lattice approximation (thin
line). Both approximations are described in the text. } 
\label{fig.avh}
\end{figure}

\begin{figure}
\caption{Number of vortex-lines per unit length in the film for $J_t=0$.
Inset: magnetization along the c-direction for $D=\lambda/4$ and $J_t=0$.} 
\label{fig.nah}
\end{figure}

\begin{figure}
\caption{Circles: vortex-line configuration for $H= 30H_{c1}$, $n_v=9$, 
$J_t=0$ and $D=\lambda$, obtained from full minimization of $E/L^2$ . 
Triangles: centered triangular lattice for $n_v=9$. The configurations
are symmetric with respect to $x_c/D=0.5$ } 
\label{fig.tri}
\end{figure}

\begin{figure}
\caption{Vortex-chain positions for $D=\lambda$ and $H=28H_{c1}$ (
$n_v=9$) .} 
\label{fig.chpj}
\end{figure}

\begin{figure}
\caption{Zero-temperature equilibrium phase diagram for $D=\lambda$.} 
\label{fig.phd}
\end{figure}

\begin{figure}
\caption{Critical currents. The arrows on the upper (lower) axis mark
the range over which $n_v$ has the value indicated for $D=\lambda/4$
($D=\lambda/2$).}  
\label{fig.jch}
\end{figure}

\end{document}